\documentclass[journal]{IEEEtran}
\usepackage{amsmath}
\usepackage{mathptmx}
\usepackage{xcolor}

\usepackage{graphicx}
\usepackage{url}
\usepackage{textcomp}
\usepackage[compress]{cite}

\begin{document}
\title{Sidebands suppression of 852 nm Cesium Faraday anomalous dispersion optical filter by argon gas}
\author{Junyu Xiong,
	Longfei Yin,
	Bin Luo,
	Jingbiao Chen
	and Hong Guo% <-this % stops a space
	
	\thanks{This work was supported by National Science Fund for Distinguished Young Scholars of China (61225003), the National Natural Science Foundation of China (61401036, 61531003), the China Postdoctoral Science Foundation (2015M580008), Youth Research and Innovation Program of BUPT (2015RC12), and the National Hi-Tech Research and Development (863) Program. 
	\emph{(Corresponding author: Hong Guo)}} %
	
	\thanks{
		J.\ Xiong, L.\ Yin, J. Chen and H. Guo are with the State Key Laboratory of Advanced Optical Communication Systems and Networks, School of Electronics Engineering and Computer Science, and Center for Quantum Information Technology, Peking University, Beijing 100871, China (email: hongguo@pku.edu.cn).
		
		B. Luo is with the State Key Laboratory of Information Photonics and Optical Communications, Beijing University of Posts and Telecommunications, Beijing 100876, China. 
		
		L. Yin is with the School of Electronic Engineering, Beijing University of Posts and Telecommunications, Beijing 100876, China.% <-this % stops a space
		}
	
	\thanks{Copyright (c) 2016 IEEE. Personal use of this material is permitted.  However, permission to use this material for any other purposes must be obtained from the IEEE by sending a request to pubs-permissions@ieee.org.} %
	}

\maketitle

\begin{abstract}
In this work, sidebands suppression induced by the buffer gas argon (Ar) in 852 nm cesium Faraday anomalous dispersion optical filter (FADOF) is investigated. FADOF performances at different Ar pressures (0 torr, 1 torr, 5 torr and 10 torr) are compared, and a single-peak transmittance spectrum with peak transmittance up to $80\%$ is achieved when the pressure of Ar reaches 5 torr. A detailed analysis shows that, this sidebands suppression comes from the depopulation enhancement by the buffer gas. This result can be generalized to other FADOFs with similar level structures such as the D2 lines of other alkali metal atoms.
\end{abstract}

\begin{IEEEkeywords}
Faraday effect,	Zeeman effect,  Optical communication
\end{IEEEkeywords}

\section{Introduction}
\IEEEPARstart{A}{s} the most commonly used atomic optical filter, Faraday anomalous dispersion optical filter (FADOF), which is based on the resonant Faraday effect, is well known for the high transmittance, the narrow bandwidth and the excellent out-of-band rejection\cite{PYeh1982,BYin1991,BYin1992}. These properties make it an important device in the fields as free-space optical communication\cite{JTang1995}, lidar\cite{APopescu2010,WHuang2009,APopescu2006}, laser frequency stabilization\cite{PWanninger1992}. In recent years, with the development of the technology and other subjects, the areas which are in demand of FADOF are observably increased, and FADOF is widely used in more fields such as ghost imaging\cite{XLiu2014}, single photon generation\cite{PSiyushev2014}, laser mode selection, which are also called Faraday laser\cite{ZTao2015,XMiao2011,ZTao2016,JKeaveney2016}, optical frequency standard\cite{WZhuang2014}, atomic spectrum recordance\cite{WKiefer2016}, magnetometer\cite{NBehbood2013}, Raman spectroscopy\cite{RAbel2009,XXue2016}, and the research of mollow triplet effect\cite{SLPortalupi2016}. For convenience in more applications and researches, a known useful tool ElecSus which could easily analyze the alkali metal D line FADOF at the extremely weak input signal is also provided\cite{ElecSus}.  

\begin{figure}[htbp]
	\centering
	\includegraphics[width=1.0\linewidth]{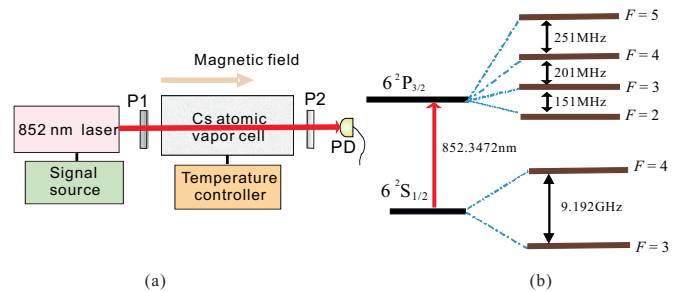}
	\caption{Experiment setup and level structure. Fig. 1(a) is the experiment setup. P1 and P2 are a pair of orthogonal polarizers, between which is the cesium vapor cell and the magnetic field. Fig. 1(b) is the level structure of 852 nm FADOF\cite{DanielASteck}. Due to the hyperfine sublevels of ${\rm{6}}{{\rm{S}}_{{\rm{1/2}}}}$ and ${\rm{6}}{{\rm{P}}_{{\rm{3/2}}}}$, the transmittance spectra of the filter exhibit multi-peak form.}
	\label{fig:false-color}
\end{figure}

FADOF is usually realized by alkali metal vapor cells\cite{PYeh1982, HGuo2010}, and will exhibits multi-peak transmittance spectra due to the hyperfine structure of alkali elements\cite{JMenders1991,RAbel2009,JZielińska2012,WKiefer2014}. However, in some specific applications such as laser frequency stabilization, Faraday laser, atomic spectrum recordance and single photon generation\cite{PWanninger1992,ZTao2015,XMiao2011,ZTao2016,JKeaveney2016,PSiyushev2014, WKiefer2016}, FADOF is used to select signal light within an extremely narrow bandwidth. In such cases, multiple peaks will introduce extra light modes and more noises, which could be vital for the system. Thus, new methods to suppress the sidebands of FADOF are significant. 
\begin{figure}[htbp]
	\centering
	\includegraphics[width=0.90\linewidth]{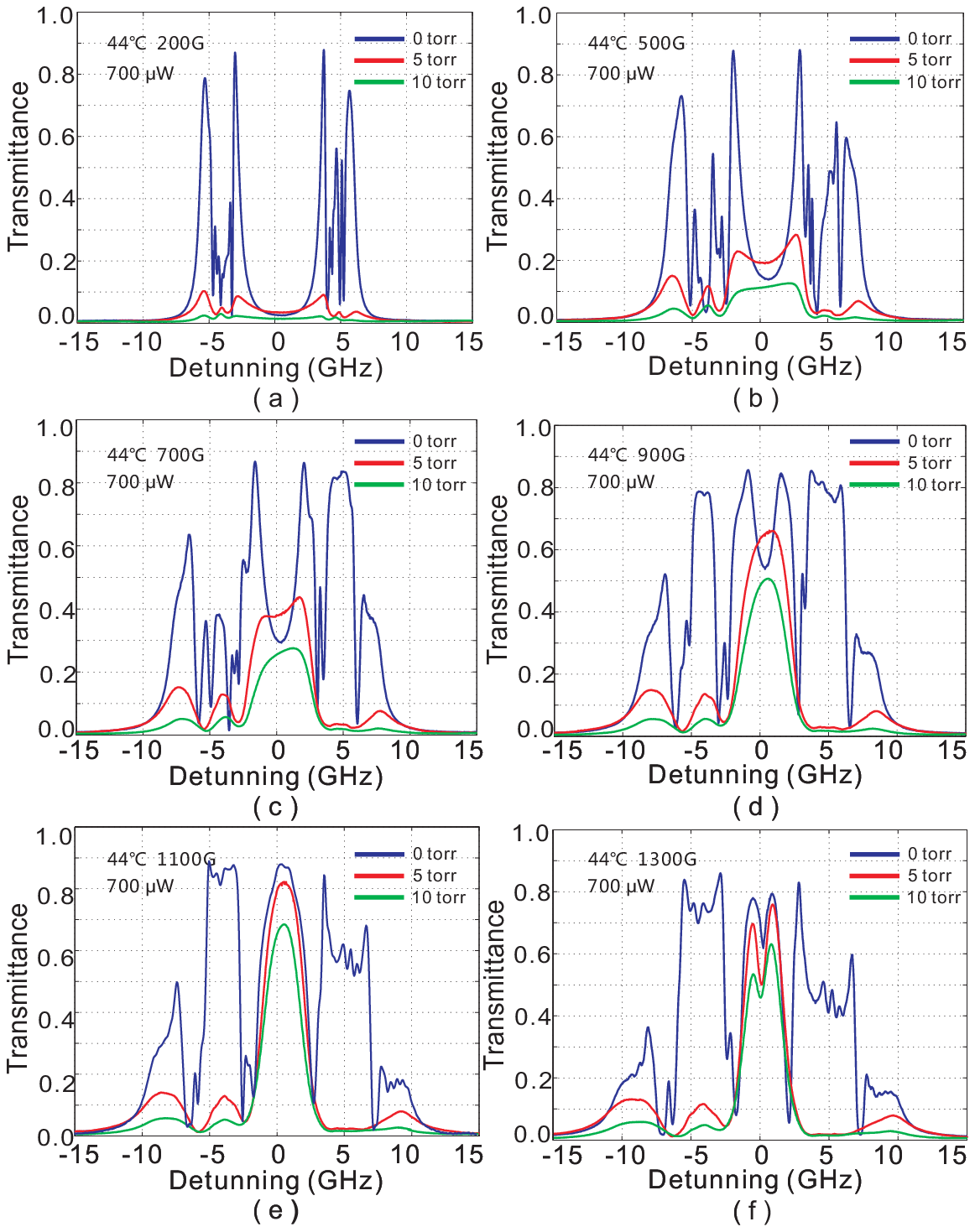}	
	\caption{The measured transmittance spectra at different magnetic field strengths, which reveal the sidebands suppression. The temperature is $44\;\,^{\circ}\mathrm{C}$. The magnetic field strengths of Fig. 2(a)--Fig. 2(f) are 200 G, 500 G, 700 G, 900 G, 1100 G and 1300 G, respectively. The laser spot diameter is 2.3 mm, and the laser power is 700 ${\mu}W$}
	\label{fig:false-color}
\end{figure}

Some efforts have been made to suppress the sidebands. Rotondaro \emph{et al}. showed that the transmittance spectrum could exhibit a narrow single peak form if the magnetic field direction and the laser path form a specific angle, while the spectrum is too sensitive to this angle\cite{MDRotondaro2015}. Another method is to use the buffer gas. Buffer gas is one of the common methods to suppress the doppler absorption and change the atomic spectrum. Inspired by this phenomenon, we tried a FADOF at 780 nm using a rubidium cell filled with buffer gas of 2 torr argon, and acquired a single peak spectrum with the peak transmittance of about $30\%$\cite{XXue2012}. Limited by the resources, this work is lack of the detailed analysis or the explanation. 

In this Letter, using cesium atomic vapor cells filled with argon of different pressures (0 torr, 1 torr, 5 torr and 10 torr), the influence of the buffer gas on the Cs 852 nm FADOF is investigated in details. Results show that the multiple peaks are efficiently suppressed by the buffer gas Ar, and the single-peak transmittance spectra can be achieved under high Ar pressures. An explanation ascribed to the enhanced depopulation effect by the buffer gas is also given. This suppression phenomenon has also been found in the cell filled with buffer gas Xe. 

\section{EXPERIMENT AND RESULTS}
The experiment setup is shown in Fig. 1. Four Cesium vapor cells (all 6 cm in length and 2 cm in width) filled with different pressures of Ar are prepared. The strength of the magnetic field and the temperature are tunable.
\begin{figure}[htbp]
	\centering
	\includegraphics[width=0.90\linewidth]{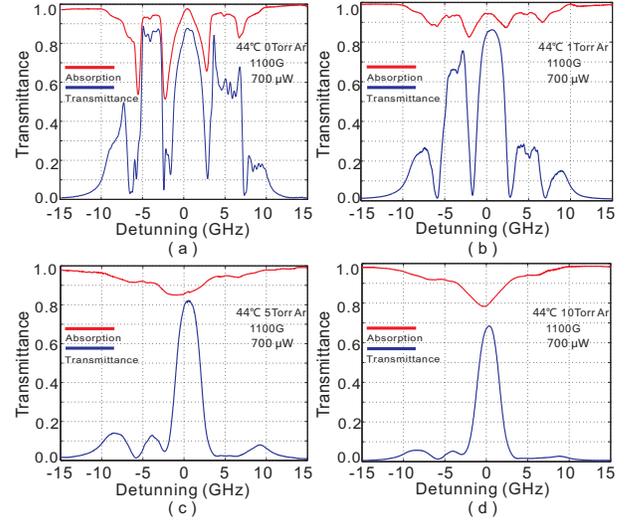}	
	\caption{The measured transmittance spectra at $44\;\,^{\circ}\mathrm{C}$ and 1100 G. The pressures of Ar in Fig. 3(a)--Fig. 3(d) are 0 torr, 1 torr, 5 torr and 10 torr, respectively. The red lines are the doppler absorption. The laser spot diameter is 2.3 mm, and the laser power is 700 ${\mu}W$}
	\label{fig:false-color}
\end{figure}

Results for the influence of buffer gas Ar on the transmittance spectra at different magnetic field strengths are shown in Fig. 2, the temperature is set to $44\;\,^{\circ}\mathrm{C}$, and the magnetic field strengths of Fig. 2(a)--Fig. 4(f) are 200 G, 500 G, 700 G, 900 G, 1100 G and 1300 G, respectively. The laser power in the experiment is 700 ${\mu}W$, and the diameter of the laser spot is 2.3 $mm$. The pressures of Ar compared in Fig. 2 are 0 torr, 5 torr and 10 torr. Due to the hyperfine sublevels of ${\rm{6}}{{\rm{S}}_{{\rm{1/2}}}}$ and ${\rm{6}}{{\rm{P}}_{{\rm{3/2}}}}$, when there is not buffer gas, the transmittance spectra of 852 nm FADOF exhibit multi-peak forms(blue lines in Fig. 2), and they become complicated as the magnetic field increases. The central peak is formed when the magnetic field is greater than 900 G, and the highest point is reached at 1100 G. When the buffer gas is added, the sidebands are observably suppressed (red and green lines in Fig. 2), while the central peak still holds the high transmittance (Fig. 2(d,e,f)). Higher Ar pressures will introduce the stronger suppression effect. 

It is obvious from Fig. 2 that the highest transmittance for all pressures of Ar is reached when the magnetic field is 1100 G. The sidebands have been obviously suppressed by the buffer gas and a high single peak remains at this point. For 5 torr Ar, the transmittance is up to $80\%$. The stronger magnetic field will decrease the peak transmittance because of the over Faraday rotation beyond 90 degree \cite{PYeh1982}. A comparison of the results at 1100 G magnetic field with different Ar pressures is shown in Fig. 3. The pressures of Ar for Fig. 3(a)--Fig. 3(d) are 0 torr, 1 torr, 5 torr and 10 torr, respectively, these four graphs clearly reveal the variation of the transmittance spectra at the influence of the buffer gas Ar. The corresponding doppler absorption signals are also obtained, which can account for the slightly decreased central peak transmittance.

The suppression effect can be evaluated by the equivalent noise bandwidth (ENBW): ${\rm{ENBW}} = {\int_{ - \infty }^{ + \infty } {T(\nu)d\nu} }/{{T_{\rm{Max}}}}$ \cite{BYin1991}, in which $\nu$ is the detunning, ${T_{\rm{Max}}}$ is the peak-transmittance, and $T(\nu)$ is the transmittance as a function of the detunning. In Fig. 3(a)--(d), ENBWs are calculated as 10.6 GHz, 7.0 GHz, 4.0 GHz, and 3.9 GHz, respectively. 

\section{Analysis of the results}
In Fig. 2, The transmittance spectra exhibit central and wing profiles under different magnetic fields. However, it is obvious that only the sidebands are significantly suppressed by the buffer gas, while the central transmittance remains. An explanation from the depopulation effect of optical pumping is given as following. 

We choose the results at $44\;\,^{\circ}\mathrm{C}$ and 1100 G to analyze the spectra. A typical pressure for Ar is 5 torr, at which the sidebands are suppressed observably and a high single-peak remains.

The transitions of cesium account for our work are shown in Fig. 4(b). The transitions marked with red arrows (A, C) in Fig. 4(b.3) are induced by the right circularly polarized light, and those marked with blue arrows (B, D) are induced by the left circularly polarized light. The relative intensities of these transitions are displayed in Fig. 4(c).

The transmittance which depends upon the optical rotation effect is proportional to the population in the lower levels\cite{PYeh1982,ETDressler1996}. The depopulation effect caused by the input light will decrease the population in the lower levels and thus restrain the optical rotation effect. As a result, the transmittance is reduced. A strong absorption of the input light will also reduce the transmittance. Hence, the high transmittance can only exhibit in the frequency range with the remarkable optical rotation effect, but without the strong depopulation effect or the strong absorption of the signal light.

For transitions A and D, atoms will be pumped from ${{m}}_{\rm{S}} = {\rm{1/2}}$ (${\rm{6}}{{\rm{S}}_{{\rm{1/2}}}}$) to ${{m}}_{\rm{S}} = {\rm{-1/2}}$ (${\rm{6}}{{\rm{S}}_{{\rm{1/2}}}}$), and from ${{m}}_{\rm{S}} = {\rm{-1/2}}$ (${\rm{6}}{{\rm{S}}_{{\rm{1/2}}}}$) to ${{m}}_{\rm{S}} = {\rm{1/2}}$ (${\rm{6}}{{\rm{S}}_{{\rm{1/2}}}}$), which will bring the depopulation effect. But for transitions B and C,
which represent ${{m}}_{\rm{S}} = {\rm{1/2}}$ (${\rm{6}}{{\rm{S}}_{{\rm{1/2}}}}$) to ${{m}}_{\rm{J}} = {\rm{3/2}}$ (${\rm{6}}{{\rm{P}}_{{\rm{3/2}}}}$) and ${{m}}_{\rm{S}} = {\rm{-1/2}}$ (${\rm{6}}{{\rm{S}}_{{\rm{1/2}}}}$) to ${{m}}_{\rm{J}} = {\rm{-3/2}}$ (${\rm{6}}{{\rm{P}}_{{\rm{3/2}}}}$), atoms can only return to the original lower levels. That is, transitions B and C cannot decrease the atoms in the lower levels and cannot bring the depopulation effect. 
\begin{figure}[htbp]
	\centering
	\includegraphics[width=0.95\linewidth]{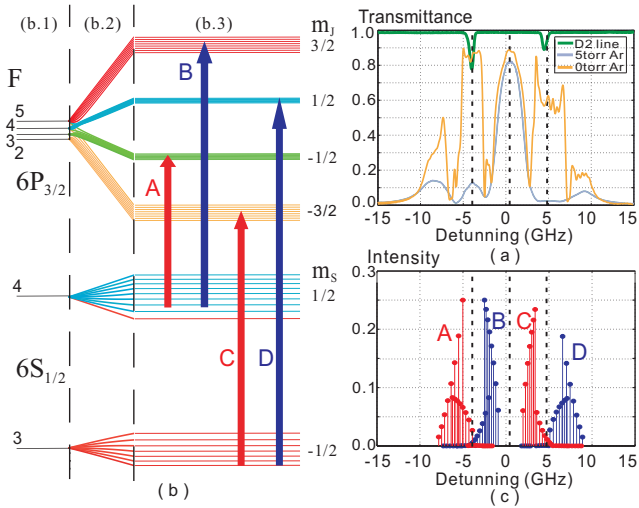}	
	\caption{Analysis diagram. Fig. 4(a) shows the transmittance spectra at $44\;\,^{\circ}\mathrm{C}$ and 1100 G, in which the pressures of Ar are 0 torr and 5torr. In Fig. 4(b), (b.1) shows the hyperfine structures of ${\rm{6}}{{\rm{S}}_{{\rm{1/2}}}}$ and ${\rm{6}}{{\rm{P}}_{{\rm{3/2}}}}$ of cesium, (b.2) is the Zeeman splitting for these two levels, and the level structures at 1100 G are in (b.3). The magnetic quantum number of electron ${\rm{m}}_{\rm{S}}$ for the red sublevels of ${\rm{6}}{{\rm{S}}_{{\rm{1/2}}}}$ is ${\rm{-1/2}}$, and for the blue sublevels is ${\rm{1/2}}$. As for ${\rm{6}}{{\rm{P}}_{{\rm{3/2}}}}$, the sublevels are marked with four colors, and the magnetic quantum numbers of electron ${\rm{m}}_{\rm{J}}$ for them are ${\rm{3/2}}$, ${\rm{1/2}}$, ${\rm{-1/2}}$ and ${\rm{-3/2}}$, respectively. The positions and relative intensities of transitions in Fig. 4(b.3) are displayed in Fig. 4(c).}
	\label{fig:false-color}
\end{figure}

When the buffer gas is introduced, it will increase the relaxation time\cite{WHapper2010}, which will weaken the relaxation effect and make the depopulation effect remarkable. When the magnetic field is weak, the sidebands arise due to the transitions A, B and C, D. But for the zero detuning, no dispersion of transitions superposes here, and cannot produce the central peak. Most atoms in the lower levels will be depopulated by transitions A and D. Thus, with the buffer gas in the cell, when the probe laser frequency is tuned to these transitions, only small amounts of atoms can populate in the lower levels due to the strong depopulation effect caused by the weak relaxation. The optical rotation effect at these frequency ranges is then suppressed, and so is the transmittance. The suppression effect goes stronger as the pressure of Ar increases. 

For transitions B and C, atoms in the lower levels cannot be depopulated, and cannot bring the suppression effect. When the magnetic field increases, and in the frequency range near the zero detuning, only the strong dispersion and weak absorption of transitions B and C superpose here, which cannot depopulate atoms in the lower levels. Thus, the strong optical rotation effect and the weak absorption make the high transmittance of the central peak.

\section{Conclusion}
The influence of the buffer gas Ar on the transmittance spectrum of 852 nm cesium FADOF is experimentally investigated. Ar buffer gases under different pressures are filled in the cesium vapor cell. The results show that, due to the depopulation effect by the buffer gas, the multi-peak sidebands are significantly suppressed. The retained single-peak is contributed by the transitions that cannot depopulate atoms from the lower levels, and the vanished sidebands are by the transitions which bring the depopulation effect. Our result can be used for other atoms with similar level structures.

\section*{Funding}
National Science Fund for Distinguished Young Scholars of China (61225003), the National Natural Science Foundation of China (61401036, 61531003), the China Postdoctoral Science Foundation (2015M580008), Youth Research and Innovation Program of BUPT (2015RC12), and the National Hi-Tech Research and Development (863) Program.

\end{document}